\documentclass[10pt,twocolumn,letterpaper]{article}

 \usepackage{iccv}              

\usepackage{booktabs}
\usepackage{multirow}
\usepackage{comment}
\usepackage{booktabs} 
\usepackage{array}    

\usepackage{ulem}
\usepackage{color}

\definecolor{iccvblue}{rgb}{0.21,0.49,0.74}
\usepackage[pagebackref,breaklinks,colorlinks,allcolors=iccvblue]{hyperref}

\def\paperID{14034} 
\def\confName{ICCV}
\def\confYear{2025}

\title{Proactive HIV Care: AI-Based Comorbidity Prediction from Routine EHR Data}

\author{
Solomon Russom \thanks{Equal contribution} 
\quad Dimitrios Kollias\footnotemark[1] 
\quad Qianni Zhang \\
Center for Multimodal AI\\
Digital Environment Research Institute\\
Queen Mary University of London \\
{\tt\small d.kollias@qmul.ac.uk}
}

\begin{document}
\maketitle
\begin{abstract}
People living with HIV face a high burden of comorbidities, yet early detection is often limited by symptom-driven screening. We evaluate the potential of AI to predict multiple comorbidities from routinely collected Electronic Health Records. Using data from 2,200 HIV-positive patients in South East London, comprising 30 laboratory markers and 7 demographic/social attributes, we compare demographic-aware models (which use both laboratory/social variables and demographic information as input) against demographic-unaware models (which exclude all demographic information). Across all methods, demographic-aware models consistently outperformed unaware counterparts. Demographic recoverability experiments revealed that gender and age can be accurately inferred from laboratory data, underscoring both the predictive value and fairness considerations of demographic features. These findings show that combining demographic and laboratory data can improve automated, multi-label comorbidity prediction in HIV care, while raising important questions about bias and interpretability in clinical AI.
\end{abstract}    

\section{Introduction}
Human Immunodeficiency Virus (HIV) continues to be a major global health concern, with millions of people living with HIV (PLWH) worldwide~\cite{unaids2023global}. Advances in antiretroviral therapy (ART) have transformed HIV into a chronic, manageable condition, enabling PLWH to live longer and healthier lives. However, increased life expectancy has also brought a growing burden of comorbidities such as cardiovascular disease, renal impairment, metabolic disorders, mental health conditions, and certain cancers~\cite{guaraldi2011noninfectious,dAD2013comorbidities}. These comorbidities often arise from a combination of factors, including chronic immune activation, ART side effects, lifestyle behaviours, and underlying demographic and social determinants of health. Early identification and proactive management of comorbidities are critical for improving patient outcomes, guiding treatment adjustments, and reducing long-term healthcare costs.


Traditionally, comorbidities in HIV care are identified via periodic clinical assessments, review of patient histories \& targeted diagnostic testing when symptoms emerge. While effective in many cases, these approaches rely heavily on clinician expertise, can be resource-intensive and may miss conditions in their subclinical stages. Laboratory markers and patient demographics are routinely available in Electronic Health Records (EHRs), but are often underutilised for systematic, proactive comorbidity screening.



AI offers powerful tools to process large-scale, heterogeneous EHR data, uncover hidden patterns, and generate timely risk predictions for multiple comorbidities simultaneously. From a \textit{technology perspective}, such systems can integrate diverse clinical and demographic features into multi-label predictive models, enabling scalable, consistent, and explainable risk stratification. From a \textit{medical perspective}, they provide an evidence-based, data-driven complement to clinical expertise, capable of identifying at-risk patients before overt symptoms appear, thus facilitating earlier interventions and personalised care. The benefits of an automated prediction system include: (i) proactive identification of high-risk patients; (ii) consistent risk assessment across large populations; (iii) the ability to capture complex, nonlinear relationships between laboratory markers, demographics, and disease risk; and (iv) improved allocation of healthcare resources.


In this paper, we develop and evaluate a range of ML and DL models for multi-label comorbidity prediction in HIV patients. Our study uses a real-world dataset of approximately 2,200 patients from South East London, spanning 2012–2023, which includes 30 routinely measured microbiology and biochemistry laboratory markers and 7 demographic and social features. The dataset also comprises disease diagnoses of these HIV patients extracted based on entries in the International Classification of Diseases (ICD-10) coding system, with hierarchical mapping from high-level categories to specific diagnostic entities.
We compare two experimental settings: a \textit{demographic-aware} approach, where demographic and social variables are included alongside laboratory data, and a \textit{demographic-unaware} approach, which uses only laboratory and social variables. This design allows us to examine not only predictive performance, but also the role and influence of demographic information in clinical risk modeling.

Our experiments show that demographic-aware models consistently outperform demographic-unaware models, with XGBoost achieving the highest overall macro F\textsubscript{1} score. The performance gain from including demographic features is particularly evident in certain comorbidity categories, underscoring their clinical relevance. We further conduct demographic recoverability experiments to quantify the extent to which sensitive attributes (e.g., gender, age, race, continent of birth) can be inferred from non-demographic inputs. Results reveal that some attributes (especially gender and age) are highly recoverable from laboratory data alone, highlighting the intertwined nature of biological markers and demographics. From both a medical and technological standpoint, these findings confirm that demographic context is valuable for accurate comorbidity prediction, while also raising important considerations for fairness and bias in AI-assisted HIV care.

\section{Related Work}

The application of AI, ML and DL, has gained substantial momentum in healthcare research over the past decade. These approaches have been widely adopted in domains such as early disease detection, prognosis estimation, clinical decision support, and resource optimization \cite{miotto2016deep,topol2019deep,shickel2018deep}. In particular, these models have demonstrated considerable effectiveness in processing structured EHR data, enabling automated risk prediction, patient stratification, and personalized treatment planning \cite{rajkomar2018scalable,bates2018big}. 

In the context of infectious diseases, several studies have explored the use of AI for improving patient outcomes. For example, AI has been employed to forecast sepsis onset from EHR data \cite{komorowski2018artificial}, predict mortality in COVID-19 patients \cite{yan2020interpretable}, and assess tuberculosis progression risks \cite{chen2022recent}. However, research specifically targeting the application of ML/DL to comorbidity prediction in HIV patients remains relatively limited.
A few recent studies have leveraged machine learning for risk modeling in HIV populations. For instance, \cite{lui2023efficient} used ML to predict cardiovascular events in HIV-positive individuals by incorporating lab results, demographics, and medication history. Similarly,  \cite{darveshwala2021chronic} applied ensemble models to identify HIV patients at risk of kidney dysfunction, demonstrating the value of EHR-derived lab features in clinical prediction tasks. However, these studies often focus on single outcomes, such as cardiovascular or renal events, and typically use smaller or less diverse input sets.
Beyond specific outcomes, other studies have explored broader comorbidity risk modeling in HIV care. One effort employed deep learning models, including Random Forest and SVM classifiers, to predict mortality risk in AIDS patients with opportunistic infections like cryptococcosis \cite{zhan2024machine}. These models achieved strong predictive performance, reinforcing the utility of AI tools in complex clinical decision-making. Furthermore, \cite{shickel2017deep} emphasized deep learning's role in EHR modeling, particularly for its ability to learn temporal and hierarchical representations, crucial for capturing the progression of chronic illnesses such as HIV.

\section{Materials and Methods}\label{section3}

\subsection{Problem Statement}

The dataset consists of $N$ data points $(\mathbf{x}_i, \mathbf{y}_i)$, $i = 1,\ldots,N$, where $\mathbf{x}_i \in \mathbb{R}^d$ represents the input features, and $\mathbf{y}_i \in \{0,1\}^m$ denotes the corresponding multi-label target vector indicating the presence or absence of $m$ comorbid conditions based on ICD-coded diagnoses. In our case, the input features $\mathbf{x}_i$ include routine laboratory test results (e.g., microbiology and biochemistry markers) and optionally demographic and social attributes.
Each $\mathbf{x}_i$ is associated with the sensitive variables $\mathbf{s}_i = \big < s_{\text{Age}} \text{ }, \text{ } s_{\text{Gender}} \text{ } , \text{ } s_{\text{Race}} , \text{ } s_{\text{Continent}}  \big>$, 
where: 
\begin{itemize}
    \item $s_{\text{Age}} \in \{\text{(21–40)}, \text{ (41–60)}, \text{ (61–80)}, \text{ (81-100)}\}$; 
    \item $s_{\text{Gender}} \in \{\text{Male}, \text{ Female}, \text{ Other}\}$;
    \item $s_{\text{Race}} \in \{\text{White}, \text{ Black}, \text{ Asian}, \text{ Mixed}, \text{ Other}\}$; and 
    \item $s_{\text{Continent}} \in \{\text{Europe}, \text{Africa}, \text{Asia}, \text{South America}, $ \\ $ \text{North America}, \text{Oceania}\}$
\end{itemize}

A sensitive variable is a label which corresponds to a protected characteristic which we do not want to base a model's decisions on. Lets us mention that we adopted a commonly accepted race and continent (of birth) classification from the U.S. Census Bureau, also aligning with those used by NHS England and UK Office for National Statistics.

To assess the influence of these attributes on model performance, we evaluate two configurations: a demographic-aware approach, which models the conditional distribution $p(\mathbf{y}_i \mid \mathbf{x}_i, \mathbf{s}_i)$ by incorporating laboratory, social and demographic features, and a demographic-unaware approach, which models $p(\mathbf{y}_i \mid \mathbf{x}_i)$ using only the laboratory and social features. This allows us to explore whether inclusion of sensitive demographic information improves predictive performance or introduces unintended biases in multi-label comorbidity prediction for HIV patients.

\subsection{Machine \& Deep Learning Models}

In this study, we evaluate the effectiveness of several ML and DL models in predicting comorbidities in HIV patients based on structured clinical and demographic data. The models include Logistic Regression (LR)\cite{menard2002applied}, Random Forest (RF)\cite{breiman2001random}, XGBoost (XGB)\cite{chen2016xgboost}, LightGBM (LGBM)\cite{ke2017lightgbm}, Multilayer Perceptron (MLP) and TabNet~\cite{arik2021tabnet}. Each model is applied to a multi-label classification setting, where the goal is to predict the presence or absence of multiple comorbidities simultaneously.

Logistic Regression (LR) is a widely used linear model for classification problems. In the multi-label context, we extend LR using the binary relevance method, where a separate LR classifier is trained for each label. LR models the log-odds of a binary outcome as a linear combination of input features, providing a probabilistic output between 0 and 1 via the logistic function. Due to its simplicity, interpretability and robustness on linearly separable data, LR serves as a strong baseline. It also offers insights into each feature's contribution via interpretable coefficients. 

Random Forest (RF) is an ensemble learning method that aggregates the predictions of multiple decision trees trained on bootstrapped samples of the data. Each tree considers a random subset of features when splitting nodes, introducing diversity and reducing the risk of overfitting. In multi-label classification, RF is extended by independently fitting one forest per label. RF is particularly suitable for structured medical data due to its robustness, scalability, and ability to model nonlinear relationships. Moreover, RF provides inherent feature importance scores, allowing for interpretation of which clinical variables are most influential in predicting specific comorbidities.

XGBoost (XGB) and LightGBM (LGBM) are gradient boosting algorithms that build additive tree models in a sequential manner. Both methods optimize for accuracy by correcting errors made by previous trees, but differ in their implementation and computational strategies. XGB employs a more regularized approach to prevent overfitting, while LGBM leverages histogram-based learning and leaf-wise growth to achieve faster training times and better performance on large datasets. In the multi-label context, both XGB and LGBM are applied using a one-vs-rest strategy. Their ability to handle missing values, capture complex interactions, and scale efficiently makes them highly effective for EHR-based prediction tasks.

Multilayer Perceptrons (MLPs) represent deep learning models capable of capturing intricate nonlinear relationships between input features and output labels. In our study, we use feedforward MLP architectures with multiple hidden layers, ReLU activations, and dropout regularization to prevent overfitting. These models are trained using backpropagation and binary cross-entropy loss for each output label. Compared to classical models, MLPs can automatically learn hierarchical feature representations, which is particularly useful when lab test values exhibit subtle patterns not easily captured by traditional methods. 

TabNet is a specialized deep learning architecture for tabular data that incorporates sequential attention mechanisms to select relevant features during training~\cite{arik2021tabnet}. Unlike traditional neural networks, TabNet performs instance-wise feature selection, enabling both improved performance and interpretability. Its ability to learn which features to attend to at each decision step allows for better generalization, especially in heterogeneous clinical data. In our multi-label setting, we apply TabNet using a multi-output wrapper, training one model per label in parallel. TabNet’s inherent explainability and strong performance on structured data make it a compelling model for clinical prediction tasks involving lab and demographic inputs.


\subsection{Dataset}

The dataset used in this study originates from South East London Hospital, comprising medical records of HIV-positive patients who have been under continuous care from 2012 to 2023. The data were sourced from two complementary Electronic Health Record (EHR) systems: the main hospital-wide EHR system and the specialty EHR used by the Sexual Health and HIV Services Department. Regular HIV care consultations are conducted at dedicated outpatient clinics within this department, where specialty EHRs are used to record detailed clinical, laboratory, and social data. In addition, patients’ interactions with other hospital specialties were also captured through the main EHR system, providing a comprehensive longitudinal view of each patient's healthcare journey.


For this study, we focused on a cohort of approximately 2200 HIV-positive patients with complete records, drawn from those actively attending HIV outpatient clinics during the specified period. The dataset includes 30 laboratory test features, primarily microbiology and biochemistry markers, as well as 7 demographic and social features. Disease diagnoses were extracted and coded based on the International Classification of Diseases (ICD) system, allowing for hierarchical classification from broad diagnostic blocks to specific disease types. This enabled a multi-label classification setup, where each patient may be associated with one or more comorbid conditions.

For our experiments, we employed a stratified k-fold cross-validation strategy, based on Iterative Stratification~\cite{sechidis2011stratification}, designed specifically for multi-label classification. It ensures that each fold maintains a representative distribution of all comorbid disease labels, preventing skewed label representation and enhancing the performance evaluation robustness across different model types.

\begin{table*}[h]
\centering
\caption{Microbiology and Biochemistry Input Features Used for Comorbidity Prediction}
\label{tab:biochem_features}
\begin{tabular}{p{3cm}|p{8cm}|p{3cm}}
\hline
\textbf{Feature} & \textbf{Description} & \textbf{Normal Range} \\
\hline
\%Hypo & Low blood sugar level & 4--6 mmol/L \\
Adjusted Calcium & Calcium level corrected for albumin & 2.2--2.6 mmol/L \\
Albumin & Protein made by the liver; marker of liver function & 3.4--5.4 g/dL \\
ALP & Enzyme linked to liver or bone conditions & 44--147 IU/L \\
AST & Enzyme elevated in liver/muscle damage & 8--33 U/L \\
Bilirubin Total & Byproduct of red blood cell breakdown & 3--20 µmol/L \\
Creatinine & Waste product indicating kidney function & 45--120 µmol/L \\
Eosinophils & Immune cells linked to allergies/infections & 0--0.4 $\times 10^9$/L \\
eGFR & Kidney filtration estimate & $>$60 mL/min \\
GGT & Liver enzyme marker of bile duct/liver damage & 5--40 U/L \\
Globulin & Immune-related proteins & 20--35 g/L \\
Hemoglobin (Hb) & Oxygen-carrying red cell protein & 12--18 g/dL \\
RNA Viral Load & Amount of HIV RNA in blood & 50--Millions \\
Lymphocytes & Key immune cells & 1000--4800/$\mu$L \\
MCH & Average hemoglobin per red cell & 27--31 pg \\
MCHC & Average hemoglobin concentration & 320--360 g/L \\
MCV & Average size of red blood cells & 80--100 fL \\
Monocytes & Immune cells against infection & 2--8\% \\
MPV & Average size of platelets & 7--9 fL \\
Neutrophils & First response immune cells & 2500--7000/$\mu$L \\
Platelets (PLT) & Blood clotting cells & 150K--450K/$\mu$L \\
Potassium & Electrolyte for nerve/muscle function & 3.6--5.2 mmol/L \\
Total Protein & Total proteins in blood & 60--83 g/L \\
Basophils & Immune cells for allergic responses & 0--1\% \\
Phosphate & Mineral for bone and energy function & 2.5--4.5 mg/dL \\
PCV & Proportion of red blood cells in blood & 35.5--48.6\% \\
RBC & Red blood cell count & 3.92--5.65 M/$\mu$L \\
RDW & Variation in red blood cell size & 12--15\% \\
Sodium & Electrolyte for fluid/nervous balance & 135--145 mEq/L \\
WBC & White blood cell count & 4.5K--11K/$\mu$L \\
\hline
\end{tabular}
\end{table*}

The input features used in this study are grouped into two categories: (A) Microbiology and Biochemistry input features, which consist of 30 continuous numerical variables measured through laboratory tests, and (B) Demographic and Social input features, which comprise 7 categorical variables describing patient background and behavioral characteristics.

\vspace{0.2cm}
\noindent
\textbf{Microbiology and Biochemistry Input Features}
The microbiology and biochemistry features represent clinical markers that are routinely measured as part of standard patient monitoring. These include indicators of liver and kidney function, immune response, and red and white blood cell activity. As detailed in Table~\ref{tab:biochem_features}, each feature is accompanied by a clinical description and its standard reference range. 
These laboratory variables are critical in HIV care, as comorbidities frequently manifest through abnormal biochemical markers even in the absence of overt symptoms. For instance, elevated liver enzymes such as AST, ALT, and GGT can signal hepatic complications, while creatinine and eGFR are key markers of renal function. Immune cell counts (including neutrophils, lymphocytes, and CD4-related indices) offer insights into immune status, which is particularly important in the context of immunocompromised individuals. These input features provide a robust clinical basis for modeling comorbidity risks and form the core of both the demographic-aware and demographic-unaware prediction pipelines.

\begin{table*}[h]
\centering
\caption{Distribution of High-Level ICD Diagnostic Categories (Comorbidities) Among HIV Patients}
\label{tab:icd_distribution}
\begin{tabular}{p{11cm}|c}
\hline
\textbf{ICD-10 Category} & \textbf{Proportion (\%)} \\
\hline
Z00–Z99: Factors influencing health status \& contact with health services & 87\% \\
A00–B99: Certain infectious and parasitic diseases                         & 46\% \\
I00–I99: Diseases of the circulatory system                                & 41\% \\
K00–K95: Diseases of the digestive system                                  & 37\% \\
E00–E89: Endocrine, nutritional and metabolic diseases                     & 29\% \\
F01–F99: Mental, behavioral and neurodevelopmental disorders               & 28\% \\
N00–N99: Diseases of the genitourinary system                              & 28\% \\
J00–J99: Diseases of the respiratory system                                & 27\% \\
M00–M99: Diseases of the musculoskeletal system \& connective tissue      & 20\% \\
D50–D89: Blood and immune-related diseases                                 & 19\% \\
G00–G99: Diseases of the nervous system                                    & 15\% \\
C00–D49: Neoplasms                                                         & 12\% \\
\hline
\end{tabular}
\end{table*}

\vspace{0.2cm}
\noindent
\textbf{Demographic and Social Input Features}
In addition to laboratory test results, the dataset incorporates seven categorical features that capture essential demographic and social factors. These include:  \textbf{Age Group} (discretized into age bands to reflect clinical risk stratification), \textbf{Gender} (aligned with self-identified categories in the EHR), \textbf{Race} (consistent with NHS classification schemes),  \textbf{Continent of Birth}, \textbf{Sexual Orientation}, \textbf{Commercial Sex Worker status} \& \textbf{Route of Infection}. These features, along with their values and distributions can be seen on Table \ref{tab:icd_distribution}.
These demographic and social variables are important as they may correlate with health disparities, behavioral risk patterns and access to care. Their inclusion enables our models' demographic-aware configuration to assess whether such features enhance predictive performance and if/how they contribute to fairness and bias in comorbidity prediction.

\vspace{0.2cm}

\noindent \textbf{Comorbidity Output Labels}
The output labels in this study correspond to high-level diagnostic categories defined by the International Classification of Diseases, 10th Revision (ICD-10). These categories are extracted from structured hospital diagnosis records and reflect the range of comorbid conditions observed among HIV patients. Each patient can be assigned to one or more diagnostic codes, making this a multi-label classification problem.
The ICD codes are grouped into 12 top-level categories, each representing a broad class of diseases/health-related conditions. As summarized in Table~\ref{tab:icd_distribution}, the most frequently observed category is Factors influencing health status and contact with health services (Z00–Z99), present in 87\% of patients. This is followed by categories for infectious diseases (A00–B99), circulatory system disorders (I00–I99), and digestive system diseases (K00–K95), each representing a significant portion of the cohort. Other prominent categories include endocrine, mental health, genitourinary, respiratory, and musculoskeletal disorders.
These comorbidity labels capture the complex and multi-systemic nature of health complications in HIV individuals. Many of these conditions are chronic or subclinical and can go undetected without structured monitoring. The inclusion of these ICD categories as output labels allows the models to learn patterns associated with disease risk across organ systems, enabling early and personalized risk assessment in clinical settings.

\subsection{Metrics}

We evaluate models' performance using metrics specifically suited to the multi-label classification nature of our comorbidity prediction task. Additionally, we employ permutation feature importance to quantify the contribution of each input feature to the models' predictive performance.

For multi-label classification problems, the standard performance metric we adopt is the macro F\textsubscript{1} score \cite{grandini2020metrics}. This metric is computed independently for each label and then averaged, treating all comorbid conditions equally regardless of their frequency in the dataset. The F\textsubscript{1} score itself is the harmonic mean of precision and recall, where precision quantifies the proportion of correctly predicted positive cases among all predicted positives, and recall reflects the proportion of true positives that were correctly identified. The F\textsubscript{1} score ranges from 0 to 1, with higher values indicating better performance.
%
%
To account for all labels (i.e., multiple comorbid conditions), we compute the macro-averaged F\textsubscript{1} score with $m$ being the number of comorbidity labels and $F_1^j$ being the F\textsubscript{1} score computed for the $j$-th label: 

\begin{equation} \label{macro_f1} \mathcal{P} = \frac{1}{m} \sum_{j=1}^{m} F_1^{j} \end{equation}
%




\subsection{Pre-Processing and Implementation Details}


In terms of pre-processing, we applied min-max normalisation \cite{singh2020investigating} to each input variable independently. In terms of implementation details, we experiment with different hyperparameters for the ML and DL models. 

In the case of RF, we used an ensemble of 10 decision trees. Each tree was grown to full depth unless all leaves became pure (based on Gini impurity) or contained one or no samples. The minimum number of samples required at a leaf node was set to one. 
For LR, we included a bias term in the decision function and applied L2 regularization to prevent overfitting. The inverse regularization strength was set to 1.0, and the optimization stopped when the tolerance for the stopping criterion reached 0.0001. LR was implemented independently for each label using the one-vs-rest strategy, a standard approach in multi-label classification tasks.
In the case of XGB and LGBM, both models were configured with default tree-based boosting parameters unless otherwise stated. These included early stopping criteria and learning rate adjustments, tailored to multi-label performance using binary relevance wrappers. 
For MLP, we implemented custom architectures consisting of multiple fully connected layers with ReLU, dropout for regularization and sigmoid output activation. MLP was trained using binary cross-entropy loss and optimized using Adam.
For TabNet, we used the PyTorch TabNet implementation with standard parameters, adjusting the number of decision steps, attention mechanisms and batch size to optimize for our specific dataset. 
%
%
All model implementations, training, and evaluations were conducted on scikit-learn and PyTorch.

\section{Experimental Results}
In the following, we present a comprehensive evaluation in which we compare the performance of all employed ML and DL models in predicting comorbidities among HIV patients. We assess each model under two settings: the demographic-aware approach, where both laboratory, demographic and social features are included as inputs, and the demographic-unaware approach, which relies solely on laboratory and social features. This comparison allows us to evaluate the contribution of demographic information to model performance. We begin by reporting overall model performance using the macro F\textsubscript{1} score, followed by a detailed analysis of results per comorbidity category. We then examine demographic recoverability to quantify the extent to which sensitive attributes can be inferred from non-demographic inputs. This structured evaluation allows us to link the observed predictive trends to both the technical properties of the models and the medical relevance of the features, providing a holistic understanding of the strengths, limitations, and practical implications of automated comorbidity prediction in HIV care. \\

\vspace{0.2cm}
\noindent
\textbf{ {\large4.1 Demographic-(Un)Aware ML/DL Models}} 

\vspace{-0.15cm}
\paragraph{Overall Performance of ML vs DL Models}
Table \ref{perf_comp} presents a performance comparison in terms of the overall $F_1$ Score (shown in \%), averaged across all folds, between the ML and DL methods, under the two studied settings. 

These results demonstrate that among all evaluated models, XGB achieved the highest overall macro $F_1$ score, establishing itself as the best-performing model in both demographic-aware and demographic-unaware approaches, while TabNet yielded the lowest performance in both approaches. This outcome aligns with expectations, as XGB is well-known for its robustness in handling imbalanced datasets, capturing complex feature interactions, and offering strong generalization, particularly in structured medical data. In contrast, TabNet, although designed for tabular data, often requires extensive hyperparameter tuning and larger datasets to realize its full potential, which may explain its underperformance in our setting.
LR and RF showed moderate results (being the second and third worst methods in both  demographic-aware vs demographic-unaware approaches). RF outperformed LR due to its capacity to model non-linearities and feature dependencies. 
LGBM also performed strongly, being the second best performing method in both demographic-aware and demographic-unaware approaches (closely matching XGB). Finally, MLP  also performed strongly, being the third best performing method in both demographic-aware and demographic-unaware approaches  Compared to tree-based methods,  MLP benefits from its ability to capture complex, non-linear interactions between features, which likely contributes to its competitive results, particularly in the demographic-aware configuration where additional categorical context can be integrated effectively. However, unlike XGB and LGBM, which are highly optimised for tabular data and robust to feature scaling, MLP is more sensitive to dataset size, feature preprocessing and hyperparameter tuning. While it substantially outperformed TabNet, its slight underperformance relative to boosting methods suggests that, in this dataset, ensemble-based learners retain a marginal edge for structured EHR data.

\begin{table}[h]
\centering
\caption{Macro $F_1$ Score (\%) of Demographic-Aware (Demo-Aware) and Demographic-Unaware (Demo-Unaware) approaches across ML and DL models}
\label{perf_comp}
\begin{tabular}{c|cc}
\hline
\textbf{Model} & \textbf{Demo-Aware} & \textbf{Demo-Unaware} \\
\hline
LR     & 37.7 & 33.9 \\
RF           & 40.7 & 38.7 \\
XGB                & \textbf{45.8} & \textbf{43.5} \\
LGBM              & 44.7 & 43.0 \\
MLP  & 44.5 & 40.8 \\
TabNet                       & 32.6 & 28.0 \\
\hline
\end{tabular}
\end{table}

\begin{table*}[h]
\centering
\caption{$F_1$ Score (\%) per comorbidity of Demographic-Aware (-A) and Demographic-Unaware (-U) approaches across ML/DL models
}

\label{perf_comp_per_label}
\scalebox{0.76}{
\begin{tabular}{l|c c c c c c c c c c c c}
\hline
\textbf{Model} & A00–B99 & C00–D49 & D50–D89 & E00–E89 & F01–F99 & G00–G99 & M00–M99 & I00–I99 & J00–J99 & K00–K95 & N00–N99 & Z00–Z99 \\
\hline
LR-A     & 56.8 & 0.0 & 38.5 & 30.2 & 37.2 & 1.1 & 7.1 & 60.5 & 34.5 & 39.2 & 54.0 & 92.8 \\
LR-U     & 56.8 & 0.0 & 37.8 & 23.3 & 18.4 & 0.0 & 3.0 & 53.6 & 31.3 & 38.4 & 51.6 & 92.8 \\
\hline
RF-A       & 60.0 & 1.5 & 42.6 & 35.7 & 34.3 & 5.0 & 11.8 & 63.1 & 39.1 & 45.3 & 56.6 & 92.8 \\
RF-U       & 58.7 & 1.0 & 42.3 & 33.1 & 25.6 & 3.6 & 10.3 & 59.7 & 37.3 & 44.2 & 55.8 & 92.8 \\
\hline
XGB-A      & 61.4 & 13.8 & 44.0 & 44.8 & 41.4 & 13.1 & 23.3 & 63.6 & 48.0 & 48.1 & 57.0 & 91.3 \\
XGB-U      & 60.3 & 9.8 & 41.9 & 41.3 & 38.1 & 11.0 & 22.5 & 61.1 & 44.7 & 46.9 & 53.2 & 91.3 \\
\hline
LGBM-A     & 61.0 & 8.6 & 42.4 & 45.4 & 43.5 & 12.7 & 20.1 & 61.9 & 45.9 & 48.9 & 54.0 & 91.4 \\
LGBM-U     & 59.5 & 8.0 & 41.1 & 43.0 & 36.4 & 10.9 & 18.9 & 60.6 & 44.3 & 48.5 & 53.4 & 91.4 \\
\hline
MLP-A      & 57.8 & 9.3 & 43.9 & 40.5 & 47.7 & 16.2 & 19.1 & 64.5 & 43.8 & 44.4 & 53.2 & 92.8 \\
MLP-U      & 55.3 & 4.7 & 42.8 & 37.2 & 33.3 & 9.7 & 16.6 & 58.8 & 43.5 & 43.7 & 50.8 & 92.8 \\
\hline
TabNet-A   & 52.2 & 0.0 & 40.8 & 21.8 & 18.2 & 3.7 & 0.0 & 59.8 & 16.5 & 38.6 & 46.6 & 92.8 \\
TabNet-U    & 43.3 & 0.0 & 39.0 & 14.7 & 8.8 & 2.8 & 0.0 & 51.2 & 13.9 & 25.0 & 45.0 & 92.8 \\
\hline
\end{tabular}
}
\end{table*}

\paragraph{Overall Performance of Demographic-Aware vs Demographic-Unaware Modeling} 

Our findings consistently show that the demographic-aware approach outperformed the demographic-unaware approach across all models. This highlights the value of incorporating demographic features into the prediction pipeline. These variables are likely to capture underlying biological factors that influence the manifestation and progression of comorbidities in HIV patients. For example, certain comorbidities may be more prevalent in specific age groups  due to behavioral, clinical, or socioeconomic patterns. By excluding these variables, the demographic-unaware models lose potentially critical context, resulting in lower performance.
This performance gap was particularly noticeable in models like LR, MLP, TabNet, where the inclusion of demographic information substantially boosted performance. All these suggest that demographic variables not only offer complementary information to clinical data but may also help the model better generalize, especially for less prevalent comorbidities. Therefore, while there may be ethical considerations in how sensitive demographic data is used, from a technical standpoint, our results demonstrate that demographic awareness enhances predictive performance in multi-label comorbidity classification for HIV patients.

\vspace{0.35cm}
\noindent
\textbf{ML vs DL
Models\ Performance per comorbidity}  
Table \ref{perf_comp_per_label} presents a $F_1$ Score (shown in \%)  comparison per  comorbidity (i.e. per class), averaged across all folds, between ML and DL methods, under the demographic-aware and demographic-unaware approaches.

Looking at the F\textsubscript{1} scores per ICD-10 block, we observe a strong alignment between model performance and label distribution. The Z00–Z99 category (e.g., routine visits or conditions not attributable to specific diseases), being the most prevalent class (87\%), was generally predicted with the highest accuracy across all models. In contrast, rare comorbidities such as neoplasms (C00–D49, 12\%) and nervous system disorders (G00–G99, 15\%) had much lower F\textsubscript{1} scores. This discrepancy is not surprising: ML and DL models often underperform on underrepresented classes due to data imbalance and a lack of sufficient patterns to learn from. 
Furthermore, for mid-prevalence conditions like mental health (F01–F99) or endocrine disorders (E00–E89), the performance varied depending on the model’s ability to leverage complex feature interactions. 
Overall, the results are both logical and expected. Models with greater capacity to model non-linear interactions and integrate demographic context performed better.

\vspace{0.35cm}
\noindent
\textbf{Demographic-Aware vs
Demographic-Unaware Modeling Performance per comorbidity} 
One can notice a consistent trend: the demographic-aware approach outperforms the demographic-unaware counterpart across nearly all ICD-10 diagnostic categories. This superiority spans a wide range of comorbidities, from common conditions (like circulatory and digestive disorders) to rarer categories (like neoplasms and nervous system diseases). These results strongly support the hypothesis that demographic context  provides valuable complementary information that aids in the accurate classification of comorbidities in HIV patients.
The only exception to this pattern is category Z00–Z99, where both approaches achieved identical performance across all models. This is a particularly interesting finding, likely attributable to two factors.

The first factor is the nature of Z00–Z99, which includes routine medical visits and general health-related encounters that are overwhelmingly frequent in the dataset (87\%). Since such visits are common across all demographic groups and not associated with disease-specific clinical or demographic patterns, the inclusion of demographic features did not offer any significant discriminative advantage. In other words, the abundance and generality of these cases make them equally easy to predict regardless of demographic context. Overall, this finding underscores the importance of tailoring model input features based on the target label. For disease categories where social determinants play a role (e.g., mental health or endocrine disorders), demographic information is highly useful. However, for generic categories such as Z00–Z99, which are driven more by healthcare utilization patterns than by demographic variation, such features offer limited added value.

The second factor is its very high distribution (87\%) in the dataset. When a class is so dominant, models are more likely to learn its patterns well, even without additional demographic context, simply because they are exposed to it frequently during training. 
By contrast, most of the other ICD-10 categories have much lower prevalence (often below 30\%), which means the model has fewer samples to learn from. In these cases, demographic features act as important side information that help the model identify hidden patterns and better compensate for the limited number of training examples. For example, mental health disorders or musculoskeletal conditions may manifest differently across age groups or vary based on gender or sexual orientation, making demographic data especially valuable in those cases.
This may suggest that if other comorbidities were as frequent as Z00–Z99, the performance gap between demographic-aware and demographic-unaware models might also shrink. However, it is also important to note that not all comorbidities are equally predictable from lab tests, regardless of distribution. So while class frequency helps, the nature of the condition and its relationship to the input features (e.g., lab values, demographics) also plays a critical role in prediction success. \\

\noindent
\textbf{ {\large4.2 Demographic Recoverability Analysis}} 

\vspace{0.2cm}

To assess the extent to which demographic attributes can be inferred from clinical variables alone, we conducted a series of demographic recoverability experiments. These experiments aimed to quantify the degree of proxy leakage, i.e., the recoverability of protected attributes from ostensibly non-demographic inputs. Such leakage is relevant for understanding potential bias risks in demographic-unaware models and for interpreting the performance gap between demographic-aware and demographic-unaware approaches.

We trained XGB to predict each demographic attribute (gender, age group, race and continent of birth) under two input configurations: (i) \textit{Lab Tests} Only, using all available laboratory and social measurements while excluding any demographic variables; and (ii)
\textit{Lab Tests and Comorbidities}, using the same laboratory and social measurements augmented with comorbidity labels, while still excluding the target demographic attribute.
For each demographic attribute, experiments were conducted in a uni-task setup (one attribute predicted at a time). Performance was evaluated using Macro-F1.
The results are summarised in Table X. 

Our demographic recoverability analysis demonstrated that lab tests encode substantial demographic information. This effect was most pronounced for gender, which was recovered with near-perfect performance, followed by age, which showed high recoverability. Race and continent of birth were partially recoverable. Adding comorbidity information to the inputs only marginally increased recoverability, suggesting that the majority of demographic signal is already embedded in lab test results.

These findings align directly with our primary observation that demographic-aware models outperform demographic-unaware models in comorbidity prediction tasks. In the absence of explicit demographic inputs, models are forced to infer demographic characteristics indirectly from noisy proxy patterns in lab data. This process is inherently imperfect: while some proxies, such as gender-related differences in hemoglobin or creatinine, are strong and reliable, others, such as ethnic variations in certain biomarkers or geographic differences in disease prevalence, are subtler and more confounded. By providing the true demographic attributes, demographic-aware models bypass this indirect inference step, allowing them to integrate precise and noise-free demographic information with lab data. This enables them to exploit well-established epidemiological associations, such as the increased cardiovascular risk in older males, the higher prevalence of certain autoimmune conditions in females, or ethnic predispositions to conditions like type 2 diabetes and sickle cell disease.

From a medical perspective, these results confirm that demographic attributes are not merely ancillary but are deeply intertwined with lab test interpretation. Many reference ranges for lab values are stratified by gender and age and population-level studies have demonstrated consistent demographic patterns in disease risk and biomarker distribution. Thus, when demographics are omitted from input, the model effectively operates with incomplete context, leading to suboptimal predictions. Conversely, incorporating accurate demographic data improves the clinical relevance of predictions by reflecting real-world diagnostic reasoning, in which clinicians routinely integrate patient demographics with lab findings to guide decision-making.

Together, these results underscore that the observed performance gap between demographic-aware and -unaware approaches is not incidental, but is a direct consequence of the demographic signal embedded in lab data and the clinical importance of demographics in disease risk stratification. They also highlight that even in demographic-unaware setups, demographic leakage is possible and should be considered in fairness and bias audits.


\begin{table}[h]
\caption{XGBoost demographic recoverability (F1 in \%) from lab tests, with and without comorbidities; CoB is Country of Birth}
\label{tab:xgb_example}
\centering
\resizebox{1\linewidth}{!}{
\begin{tabular}{l|cccc}
\hline
XGB & Gender & Age & Race & CoB \\
\hline
Lab Tests & 92.8 & 45.4 & 35.9  & 27.3  \\
Lab Tests \& Comorbidities & 92.8 & 46.9 & 36.1 & 27.5 \\
\hline
\end{tabular}
}
\end{table}


\section{Conclusions}
We evaluated AI's potential to predict multiple comorbidities from routinely collected EHR of people living with HIV. We further examined the impact of demographic awareness on such predictive modelling. Through extensive ML and DL experiments, we demonstrated that models incorporating demographic information consistently outperformed their demographic-unaware counterparts. 
Our findings underscore two key implications. First, demographic variables, when available and used responsibly, can provide meaningful context that improves predictive performance. Second, the relationship between laboratory measurements, comorbidities and demographics is complex and interdependent, suggesting that future research should explore richer multimodal and longitudinal data integration. At the same time, these results call for careful consideration of fairness, bias mitigation and privacy when incorporating sensitive demographic attributes into model development.

\section{Acknowledgments}
The work of Solomon Russom had been supported by Kings College Hospital NHS Trust.

{
    \small
    \bibliographystyle{ieeenat_fullname}
    \bibliography{main}
}

\end{document}